\documentclass[%
 aip,
 sd,%
 amsmath,amssymb,
 reprint,%
]{revtex4-1}

\usepackage{subfigure}
\usepackage{rotating}
\usepackage{listings}
\usepackage{caption}
\usepackage{amsmath}
\usepackage{url}
\usepackage{graphicx}
\usepackage{dcolumn}
\usepackage{bm}

\usepackage{soul,color}

\definecolor{orange}{rgb}{1.0,0.4,0.0}

\newcommand{\beq}{\begin{equation}}
\newcommand{\eeq}{\end{equation}}
\newcommand{\bseq}{\begin{subequations}}
\newcommand{\eseq}{\end{subequations}}
\newcommand{\beqn}{\begin{eqnarray}}
\newcommand{\eeqn}{\end{eqnarray}}
\newcommand{\ba}{\begin{array}}
\newcommand{\ea}{\end{array}}
\newcommand{\bct}{\begin{center}}
\newcommand{\ect}{\end{center}}
\newcommand{\btmz}{\begin{itemize}}
\newcommand{\etmz}{\end{itemize}}
\newcommand{\benum}{\begin{enumerate}}
\newcommand{\eenum}{\end{enumerate}}

\newcommand{\norm}[1]{\| #1 \|}                 

\newcommand{\be}{\begin{equation}}
\newcommand{\ee}{\end{equation}}

\newcommand{\cplxs}{ C\kern -.35em \rule{0.03 em}{.7 ex}~   }

\def\complex{\hbox{C\kern -.45em \rule{0.03 em}{1.5 ex}}~}

\newcommand{\bi}{\begin{itemize}}
\newcommand{\ei}{\end{itemize}}

\renewcommand{\BibitemShut}[1]{}

\newcommand{\non}{\nonumber}
\newcommand{\ds}{\displaystyle}

\newcommand{\mrd}{\mathrm{d}}
\newcommand{\mre}{\mathrm{e}}
\newcommand{\mri}{\mathrm{i}}

\newcommand{\fvec}{{\bf f}}

\newcommand{\bu}{{\bf u}}
\newcommand{\bU}{{\bf U}}

\newcommand{\bphi}{\mbox{\boldmath$\phi$}}
\newcommand{\bpsi}{\mbox{\boldmath$\psi$}}

\newcommand{\bkappa}{\mbox{\boldmath$\kappa$}}

\newcommand{\p}{\partial}

\providecommand{\norm}[1]{\lVert#1\rVert}

\newcommand{\moa}[1]{{{\color{black}#1}}}

\DeclareMathOperator*{\minimize}{minimize}
\DeclareMathOperator*{\subject}{subject~to}

\begin{document}

\preprint{AIP/123-QED}

\title[A low-order decomposition of turbulent channel flow]
{A low-order decomposition of turbulent channel flow via resolvent analysis and convex optimization}

\author{R. Moarref}
 \affiliation{Graduate Aerospace Laboratories, California Institute of Technology, Pasadena, CA 91125, USA}
\author{M. R. Jovanovi{\'c}}
 \affiliation{Electrical and Computer Engineering, University of Minnesota, Minneapolis, MN 55455, USA}
\author{J. A. Tropp}
 \affiliation{Computing \& Mathematical Sciences, California Institute of Technology, Pasadena, CA 91125, USA}
\author{A. S. Sharma}
 \affiliation{Engineering and the Environment, University of Southampton, Southampton, SO17 1BJ, UK}
\author{B. J. McKeon}
 \affiliation{Graduate Aerospace Laboratories, California Institute of Technology, Pasadena, CA 91125, USA}

\date{\today}

\begin{abstract}

We combine resolvent-mode decomposition with techniques from convex optimization to optimally approximate velocity spectra in a turbulent channel. The velocity is expressed as a weighted sum of resolvent modes that are dynamically significant, non-empirical, and scalable with Reynolds number. To optimally represent DNS data at friction Reynolds number $2003$, we determine the weights of resolvent modes as the solution of a convex optimization problem. Using only $12$ modes per wall-parallel wavenumber pair and temporal frequency, we obtain close agreement with DNS-spectra, reducing the wall-normal and temporal resolutions used in the simulation by three orders of magnitude.

\end{abstract}

\maketitle

Wall-bounded turbulent flows are dominated by coherent structures, see e.g. Smits \emph{et al.}\cite{smimckmar11}, which motivates the search for their low-order decomposition and modeling. Ideally, a low-order model captures the essential flow physics and is amenable to the techniques that are advanced in systems control and optimization theories. Therefore, it offers several advantages for understanding and controlling wall-turbulence. While a low-order decomposition focuses on capturing the relevant flow quantities such as the velocity fluctuations, a low-order model is concerned with explaining and predicting the flow behavior using the low-order decomposition. \moa{Most low-order decompositions are driven by experimental or simulation data and their empirical nature may obscure important flow dynamics~\cite{berhollum93,row05,sch10,tum11,mez13}.} 

In this paper, we show that a gain-based low-order decomposition that is obtained from the Navier-Stokes equations (NSE) can be used to approximate the turbulent velocity spectra. Recent developments by McKeon, Sharma, and co-workers~\cite{mcksha10,mckshajac13,shamck13,moashatromckJFM13} have highlighted the power of this decomposition in capturing several features of wall-turbulence and their Reynolds-number scalings. The proposed decomposition~\cite{mcksha10}, discussed later, expresses the velocity as a weighted sum of resolvent modes and exhibits two important advantages relative to the other low-order decompositions: (i) The resolvent modes are non-empirical since they represent the most amplified shapes by the linear mechanisms in the NSE; and (ii) The Reynolds-number scaling of the resolvent modes are known~\cite{moashatromckJFM13}. These properties are essential to predicting the behavior of wall-turbulence at high Reynolds numbers. The remaining challenge is related to computation and scaling of the weights that determine the contribution of the resolvent modes to the turbulent kinetic energy and represent the nonlinear interaction of the resolvent modes. Here, we compute the weights such that the resolvent-mode decomposition optimally matches the two-dimensional velocity spectra from direct numerical simulations (DNS) of Hoyas \& Jimenez~\cite{hoyjim06} for channel flow with $Re_\tau = 2003$. Even though this yields an empirical way for computing the weights, theoretical developments for their non-empirical determination is the subject of ongoing research.

Consider the NSE
	\be
	\ba{l}
	\partial \bu/ \partial t
	\, + \,
	(\bu \cdot \nabla) \bu
	\, + \,
	\nabla P
	\; = \;
	(1/Re_\tau)  \nabla^2 \bu,
	~~
	\nabla \cdot \bu
	\; = \;
	0,
	\ea
	\label{eq.NS}
	\ee
where $\bu = [\,u~v~w\,]^T$ is the velocity vector in the streamwise $x$, wall-normal $y \in [0,2]$, and spanwise $z$ directions, $t$ is time, $P$ is the pressure, and $\nabla$ is the gradient operator. The Reynolds number $Re_\tau = u_\tau h/\nu$ is defined based on the channel half-height $h$, kinematic viscosity $\nu$, and friction velocity $u_\tau = \sqrt{\tau_w/\rho}$, where $\tau_w$ is the shear stress at the wall, and $\rho$ is the density. Velocity is normalized by $u_\tau$, spatial variables by $h$, time by $h/u_\tau$, pressure by $\rho u_\tau^2$, and plus denotes normalization by the viscous scale, e.g. $y^+ = Re_\tau y$. The resolvent-mode decomposition for channel flow is summarized in equations~(\ref{eq.Fourier}) and~(\ref{eq.svd-u}) below, see also Fig.~\ref{fig.block-diag}. The Fourier modes are the appropriate basis in the homogeneous wall-parallel directions and time,
	\be
	\bu (x,y,z,t)
	\; = \;
	\ds{
	\iiint_{-\infty}^{\infty}
	}
	\,
	\hat{\bu} (y, \bkappa, \omega)\,
	\mre^{\mri
	(
	\kappa_x x
	\, + \,
	\kappa_z z
	\, - \,
	\omega t
	)
	}
	\mrd \kappa_x \,
	\mrd \kappa_z \,
	\mrd \omega.
	\label{eq.Fourier}
	\ee
Here, the hat denotes the Fourier coefficients, $\kappa_x$ and $\kappa_z$ are the streamwise and spanwise wavenumbers, and $\omega$ is the temporal frequency. The nonlinear term in~(\ref{eq.NS}) is considered as a forcing term $\fvec = -(\bu \cdot \nabla) \bu$ that drives the velocity fluctuations around the turbulent mean velocity $\bU = [\,U(y)~0~0\,]^T = \hat{\bu}(y,0,0,0)$,
\moa{
	\be
	-\mri \omega \hat{\bu}
	\, + \,
	(\bU \cdot \nabla) \hat{\bu}
	\, + \,
	(\hat{\bu} \cdot \nabla) \bU
	\, + \,
	\nabla \hat{p}
	\, - \,
	(1/{Re}_\tau) \Delta \hat{\bu}
	\; = \;
	\hat{\fvec},
	~~
	\nabla \cdot \hat{\bu}
	\; = \;
	0,
	\non
	\ee
where $\nabla = [\,\mri \kappa_x~\p_y~\mri \kappa_z\,]^T$ and $\Delta = \p_{yy} - \kappa_x^2 - \kappa_z^2$.
}
The input-output relationship between $\hat{\fvec}$ and $\hat{\bu}$ is governed by the resolvent operator $H$, see Moarref \emph{et al.}\cite{moashatromckJFM13} for details,
	\be
	\hat{\bu} (y, \bkappa, \omega)
	\, = \,
	H (\bkappa, \omega) \, \hat{\fvec} (y, \bkappa, \omega),
	\label{eq.H}
	\non
	\ee 
where $\bkappa = [\,\kappa_x~\kappa_z\,]$ is the wavenumber vector. For any $(\bkappa, \omega)$, a complete basis in $y$ is determined using the Schmidt (singular value) decomposition of $H$. This yields two orthonormal sets of unit-energy forcing modes $\hat{\bphi}_j = [\,\hat{f}_{1j}~\hat{f}_{2j}~\hat{f}_{3j}\,]^T$ and unit-energy response modes $\hat{\bpsi}_j = [\,\hat{u}_j~\hat{v}_j~\hat{w}_j\,]^T$ (henceforth ``resolvent modes") that are ordered by the corresponding gains $\sigma_1 \geq \sigma_2 \geq \cdots > 0$. Each Fourier coefficient in~(\ref{eq.Fourier}) can be approximated using a weighted sum of the first $N$ resolvent modes
	\be
	\ba{rcl}
	\hat{\bu} (y, \bkappa,\omega)
	&\!\! = \!\!&
	\ds{\sum_{j = 1}^{N}}
	\;
	\chi_j(\bkappa,\omega) 
	\,
	\sigma_j(\bkappa,\omega) 
	\,
	\hat{\bpsi}_j (y, \bkappa,\omega),
	\ea
	\label{eq.svd-u}
	\ee
where the weights $\chi_j$ represent the projection of $\hat{\fvec}$ onto the first $N$ forcing modes $\hat{\bphi}_j$
	\be
	\hat{\fvec} (y, \bkappa,\omega)
	\, = \,
	\ds{\sum_{j = 1}^{N}}
	\;
	\chi_j(\bkappa,\omega) 
	\,
	\hat{\bphi}_j (y, \bkappa,\omega),
	~~
	\chi_j (\bkappa,\omega)
	\, = \,
	\ds{\int_{0}^{2}}
	\hat{\bphi}_j^* (y, \bkappa,\omega)
	\,
	\hat{\fvec} (y, \bkappa,\omega)
	\,
	\mrd y.
	\label{eq.svd-f-chi}
	\non
	\ee 
\moa{Even though the energy of each mode is determined by the product of the corresponding gain and weight, we separately study $\sigma_j$ and $\chi_j$ to distinguish the linear and nonlinear mechanisms in the NSE.}

Each resolvent mode represents a propagating wave with streamwise and spanwise wavelengths $\lambda_x = 2\pi/\kappa_x$ and $\lambda_z = 2\pi/\kappa_z$ and streamwise speed ${c} = \omega/\kappa_x$, as suggested by the Fourier decomposition~(\ref{eq.Fourier}). The resolvent modes are localized around the critical wall-normal location $y_c$ where the mode speed equals the local mean velocity~\cite{mcksha10}, i.e. $c = U(y_c)$. Moarref \emph{et al.}\cite{moashatromckJFM13} analytically established that the Reynolds number scalings of the resolvent modes are determined by the mode speed and the different regions of the mean velocity, i.e. the inner- and outer-scaled regions and the logarithmic overlap region in the classical picture~\cite{col56}. Owing to the integral role of $c$ in the wall-normal localization and scalings of the resolvent modes, the Fourier-transformed variables are parameterized by $c$ instead of $\omega$ in the rest of this paper. In addition, we confine our attention to the modes with $0 \leq c \leq U_{cl}$ where $U_{cl} = U(1)$ denotes the centerline velocity. This conservative choice is motivated by a range of observations, summarized by LeHew \emph{et al.}\cite{lehguamck11}, that the convective velocity of energetic eddies in turbulent flows is approximately confined between $8 u_\tau$ and $U_{cl}$. 

    \begin{figure}
    \begin{center}
    \includegraphics[width=0.55\columnwidth]
    {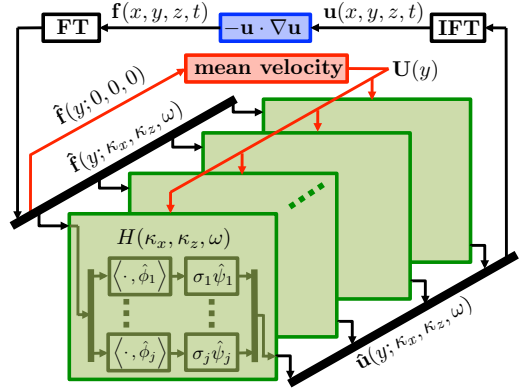}
    \vspace{-0.5cm}
    \end{center}
    \captionsetup{justification=raggedright}
    \caption{A block-diagram illustrating the linear mechanisms in the NSE and the nonlinear feedback that drives the fluctuations and sustains the mean velocity. FT and IFT stand for Fourier transform and inverse Fourier transform. The resolvent operator $H$ governs the relationship between the nonlinearity $\hat{\fvec}$ and the fluctuations $\hat{\bu}$. The singular value decomposition of $H$ yields an ordered set of most amplified forcing and response modes that are used as a basis for wall-normal decomposition of the fluctuations.}
    \label{fig.block-diag}
    \end{figure}

The effectiveness of the resolvent-mode decomposition for representing the turbulent spectra is evaluated by finding the optimal weights that minimize the deviation between the spectra resulting from DNS~\cite{hoyjim06} and from the resolvent-mode decomposition. The premultiplied three-dimensional streamwise energy spectrum is defined as
	\be
	\ba{rcl}
	E_{uu} (y, \bkappa, c)
	&\!\! = \!\!&
	\kappa_x^2 \kappa_z\,
	\hat{u} (y, \bkappa, c)
	\;
	\hat{u} (y, \bkappa, c)^*.
	\ea
	\label{eq.Euu-u}
	\ee
The additional power of $\kappa_x$ in~(\ref{eq.Euu-u}) facilitates computation of the time-averaged premultiplied two-dimensional streamwise energy spectrum by integration over $c$ instead of $\omega$
	\be
	\hskip-2.62cm
	E_{uu} (y,\bkappa)
	\; = \;
	\ds{
	\int_{0}^{U_{cl}}
	}
	\,
	E_{uu} (y, \bkappa, c)
	\,
	\mrd c.
	\label{eq.Euu-ykxkz}
	\ee
The contribution of the first $N$ resolvent modes to $E_{uu} (y, \bkappa, c)$ is determined by substituting $\hat{u}$ from~(\ref{eq.svd-u}) in~(\ref{eq.Euu-u}). Since $E_{uu}$ is a quadratic function of the resolvent weights, selecting $\chi_j$ in order to minimize the deviation from the simulation-based spectra results in a non-convex optimization problem. The globally optimal solution of this class of problems is known to be difficult to find. To overcome this challenge, we introduce an $N \times N$ weight matrix $X (\bkappa, c)$ whose $ij$-th element is determined by
	\be
	X_{ij} (\bkappa, c)
	\; = \;
	\chi_i^* (\bkappa, c)
	\,
	\chi_j (\bkappa, c),
	\label{eq.X-ij}
	\ee
and express $E_{uu}$ as a linear function of $X$
	\be
	\ba{rcl}
	E_{uu} (y, \bkappa, c)
	&\!\! = \!\!&
	\mbox{Re} 
	\Big\{
	\mbox{tr}
	\big(
	A_{uu} (y, \bkappa, c)
	\;
	X (\bkappa, c) 
	\big)
	\Big\}.
	\ea
	\label{eq.Euu-trace}
	\ee
Here, $\mbox{Re}$ is the real part of a complex number, $\mbox{tr}(\cdot)$ is the matrix trace, $A_{uu} (y, \bkappa, c)$ is the $N \times N$ energy density matrix whose $ij$-th element is determined by the resolvent modes,
	\be
	A_{uu,ij} (y, \bkappa, c)
	\; = \;
	\kappa_x^2 \kappa_z \,
	\,
	\sigma_i(\bkappa, c) \,
	\sigma_j(\bkappa, c) \,
	\hat{u}_i (y, \bkappa, c) \, \hat{u}_j (y, \bkappa, c)^*.
	\label{eq.Euu-tilde-ij}
	\ee
The expressions~(\ref{eq.Euu-ykxkz})-(\ref{eq.Euu-tilde-ij}) for the wall-normal and spanwise energy intensities $E_{vv}$ and $E_{ww}$ and the Reynolds stress $E_{uv}$ are obtained similarly.   

For given $\bkappa$, we formulate the following optimization problem
	\be
	\ba{rl}
	\minimize\limits_{X, \, e}
	&
	e (\bkappa)
	\\[0.08cm]
	\subject
	&
	\moa{
	\dfrac{
	\norm{
	E_{r,\mathrm{DNS}} (y,\bkappa)
	\, - \,
	\int_{0}^{U_{cl}}
	\,
	\mbox{Re} 
	\big\{
	\mbox{tr}
	\big(
	A_{r} (y, \bkappa, c)
	\;
	X (\bkappa, c) 
	\big)
	\big\}
	\,
	\mrd c
	}^2
	}
	{
	\norm{E_{r,\mathrm{DNS}} (y,\bkappa)}^2
	}
	\; \leq \; 
	e (\bkappa)
	}
	\\[0.15cm]
	&
	X(\bkappa,c)
	\succeq 
	0
	\\[0.15cm]
	&
	\mbox{rank}\big(X(\bkappa,c)\big)
	\; = \;
	1,
	\ea
	\label{eq.opt}
	\ee
for the weight matrices $X(\bkappa,c)$ and the deviation error $e (\bkappa)$. Here, the DNS-based spectra $E_{r,\mathrm{DNS}}$ and the energy density matrices $A_r$, with 
	$
	r
	= 
	\{ uu, vv, ww, uv \},	
	$
are the problem data (obtained from simulations\cite{hoyjim06} and the resolvent modes, respectively). The optimization problem~(\ref{eq.opt}) is formulated in order to simultaneously minimize the deviation errors for all three velocity spectra and the Reynolds stress co-spectrum. In the first constraint, the integral term quantifies the model-based energy spectrum and is obtained from~(\ref{eq.Euu-ykxkz}) and~(\ref{eq.Euu-trace}). The last two constraints follow from the definition of the weight matrix $X$ in~(\ref{eq.X-ij}), and require it to be positive semi-definite and rank-1. The norm $\moa{\norm{g}^2 = \int_{y^+ = 5}^{y = 1} |g (\ln y^+)|^2 \mrd \ln y^+}$ is defined such that the deviation between the DNS-based and the model-based spectra is equally penalized in the channel core and close to the walls. This norm is different from the standard energy norm ($L_2$) which is used to compute the resolvent modes. The lower limit $y^+ = 5$ equals the smallest wall-normal location where the DNS data is available. 

For any $\bkappa$,~(\ref{eq.opt}) is discretized with $N_y$ logarithmically-spaced points between $y^+ = 5$ and $y = 1$ and $N_c$ linearly-spaced points between $c = 0$ and $U_{cl}$. Furthermore, by defining
	\[
	X_l \; = \; X(\bkappa,c_l), 
	~~
	A_{r,lm} \; = \; A_{r}(y_m,\bkappa,c_l),
	~~
	l \; = \; \{1, 2, \ldots, N_c\}, 
	~
	m \; = \; \{1, 2, \ldots, N_y\},
	\] 
we can use~(\ref{eq.Euu-ykxkz}) and~(\ref{eq.Euu-trace}) to obtain
	\be
	E_{r} (y_m, \bkappa)
	\; = \;
	\ds{\sum_{l \, = \, 1}^{N_c}}
	\;
	\mbox{Re} 
	\big\{
	\mbox{tr}
	(
	A_{r,lm}
	\;
	X_l
	)
	\big\}
	,  
	~~
	m 
	\, = \,
	\{1,2,\ldots,N_y\}
	,
	~~
	r
	\, = \,
	\{uu, vv, ww, uv\}. 
	\non
	\ee
The DNS data are interpolated on the wall-parallel wavenumbers and the wall-normal locations that are considered in the minimization problem. 

We note that the rank constraint represents the only source of non-convexity in the optimization problem~(\ref{eq.opt}). For the special case where $N = 1$, this problem is convex since the rank constraint is eliminated. Even though one resolvent mode per $\bkappa$ and $c$ is sufficient to represent and predict the streamwise energy intensity at high $Re_\tau$~\cite{moashatromckJFM13}, it cannot simultaneously represent the turbulent velocity spectra in the streamwise, wall-normal, and spanwise directions~\cite{moashatromck13-AIAA}. Therefore, we consider the general case where $N > 1$. The computational challenge is that the optimization problem~(\ref{eq.opt}) is not convex for $N > 1$ due to the rank constraint on $X_l$. In general, there is no guarantee that the globally optimal solution of non-convex problems can be found. However, a rank-1 solution of problem~(\ref{eq.opt}) that yields the globally optimal deviation error $e (\bkappa)$ can be obtained using the following procedure, see Huang \& Palomar\cite{huapal10} for details: 
	\bi
	
	\item[(i)] Remove the rank constraint in~(\ref{eq.opt}) and solve the resulting semi-definite programming problem using convex optimization solvers such as CVX~\cite{cvx}. This yields a globally optimal deviation error $e$ and a typically full-rank optimal solution $\{X_l\}_{l = 1,2,\ldots,N_c}$. 
	\vspace{-0.14cm}
	\item[(ii)] While $\sum_{l = 1}^{N_c} \mbox{rank}(X_l)^2 > N_c$, iterate (iii)-(vi).
	\vspace{-0.14cm}
	\item[(iii)] Decompose $X_l = V_l V_l^*$ for $l = 1, 2, \ldots, N_c$.
	\vspace{-0.14cm}
	\item[(iv)] Find a non-zero Hermitian solution $\{Y_l\}_{l = 1,2,\ldots,N_c}$, with the same rank as $\{X_l\}_{l = 1,2,\ldots,N_c}$, to the following linear system of equations
	\be
	\ds{\sum_{l = 1}^{N_c}}
	\;
	\mbox{Re}\Big\{
	{\mbox{tr}\big(V_l^* \, A_{r,lm} \, V_l \, Y_l \big)}
	\Big\}
	\; = \;
	0
	,~~
	m 
	\, = \,
	\{1,2,\ldots,N_y\}
	,
	~~
	r
	\, = \,
	\{uu, vv, ww, uv\}.
	\non
	\ee 
	\vspace{-0.14cm}
	\item[(v)] Let $\lambda$ be the maximum of the absolute values of the eigenvalues of $Y_1$ to $Y_{N_c}$.
	\vspace{-0.14cm}
	\item[(vi)] Update $X_l = V_l (I_l - Y_l/\lambda) V_l^*$, where $I_l$ is the identity matrix of the same size as $Y_l$.
	\ei
	
The above procedure begins with the globally optimal solution in the absence of the rank constraint and, at each iteration, computes a new solution with a smaller rank without changing the globally optimal deviation error $e$. The update law in step (vi) involves two terms: The term $V_l V_l^*$ equals the present weight matrix, cf.~step (iii), and the term $V_l (Y_l/\lambda) V_l^*$ lies in the null space of the operator that maps the weight matrix to the energy spectra, cf. step (iv) and equation~(\ref{eq.Euu-trace}). Consequently, the energy spectra resulting from the updated $X_l$ is the same as the original globally optimal solution, and the deviation error remains unchanged. In addition, since $\lambda$ is an eigenvalue of $Y_l$, subtracting $V_l (Y_l/\lambda) V_l^*$ from $V_l V_l^*$ reduces the rank of $X_l$. Notice that the rank of the globally optimal solution can be reduced as long as the linear system of equations in (iv) has a non-zero solution~\cite{huapal10}. This system of equations consists of $4N_y$ equations and $N_c N^2$ unknowns. Therefore, a non-zero solution exists if $4N_y < N_c N^2$. To satisfy this requirement, we choose $N_y = 60$, $N_c = 100$, and $N > 1$. It should be noted that computation of the optimal solution becomes more expensive for larger values of $N_y$, $N_c$, and $N$. This is because a larger $N_y$ results in a larger number of constraints and larger values of $N_c$ and $N$ result in a larger number of unknowns in the problem. In addition, a larger $N$ increases the number of wall-normal grid points that are required to capture the complex shapes of higher-order resolvent modes. Our results show negligible sensitivity to additional increase in $N_y$ and $N_c$.

    \begin{figure}
    \begin{center}
    \begin{tabular}{ccc}
    \subfigure{\includegraphics[width=0.33\columnwidth]
    {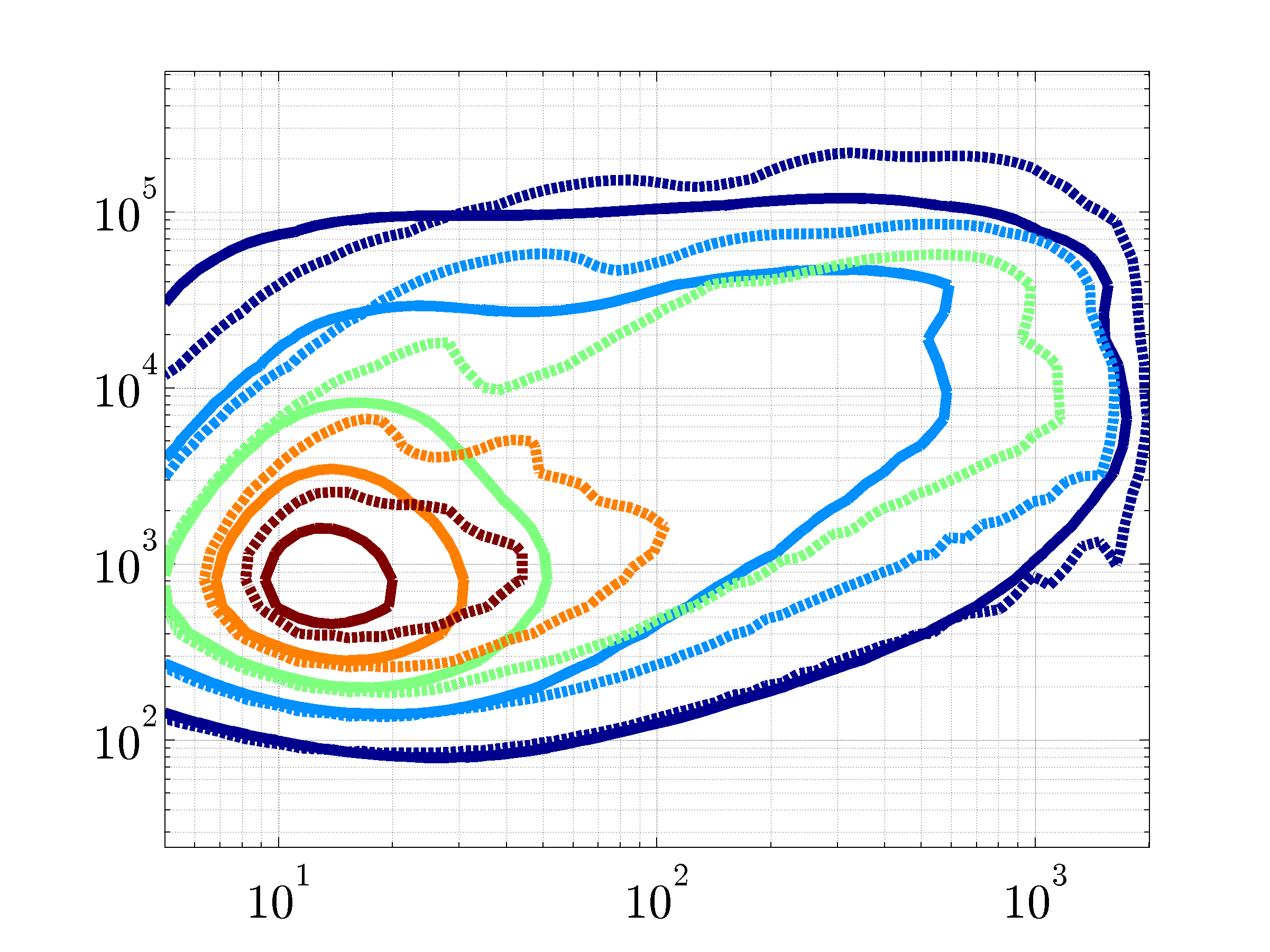}
    \label{fig.Euu_vs_yp_lxp_DNSb_Wr_R2003_match_uvw_uvr_y60_c0_100_Uc_s1_2_3_072613_c}}
    &
    \hskip-0.5cm
    \subfigure{\includegraphics[width=0.33\columnwidth]
    {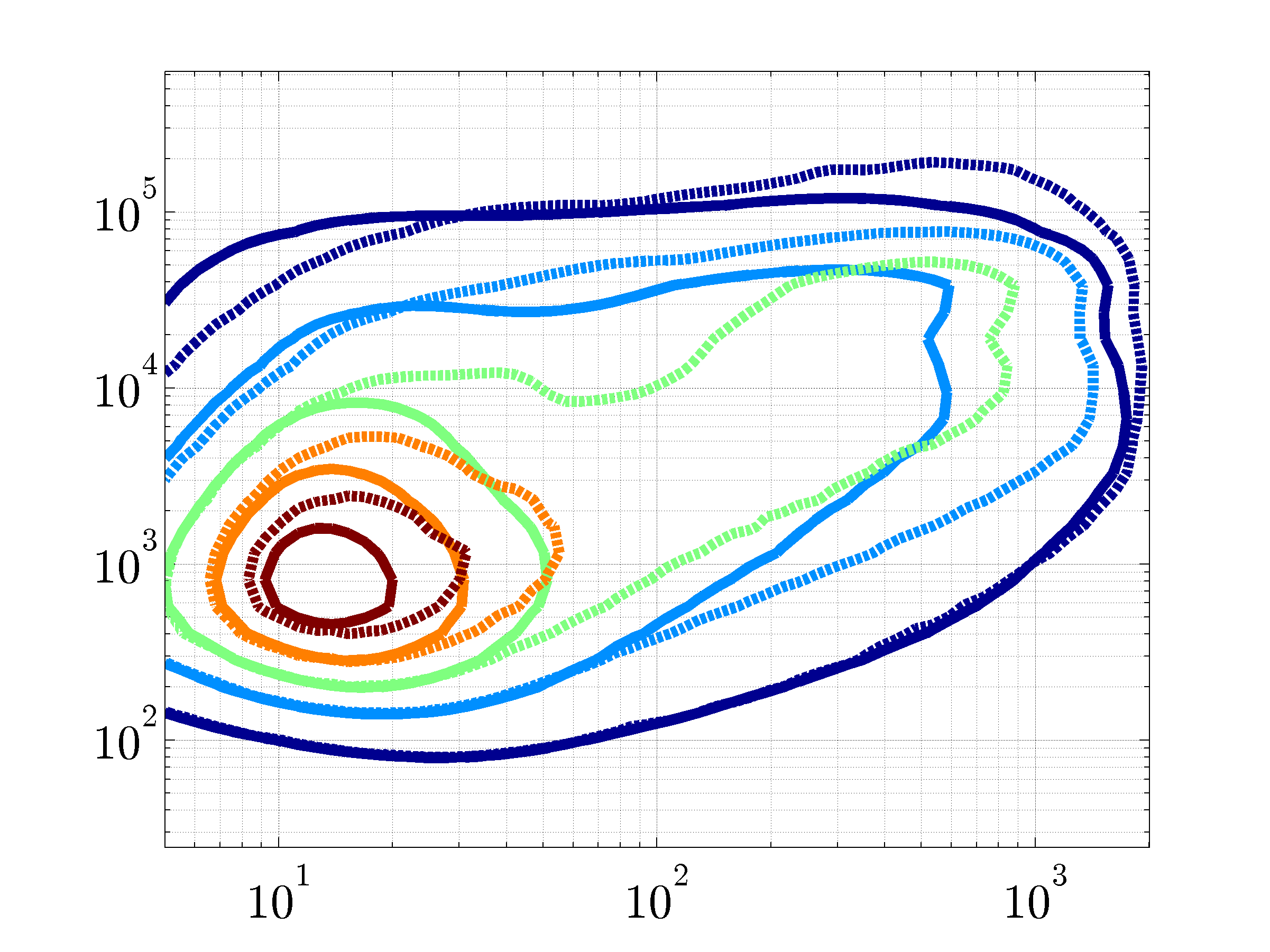}
    \label{fig.Euu_vs_yp_lxp_DNSb_Wr_R2003_match_uvw_uvr_y60_c0_100_Uc_s1_2_11_072613_c}}
    &
    \hskip-0.5cm
    \subfigure{\includegraphics[width=0.33\columnwidth]
    {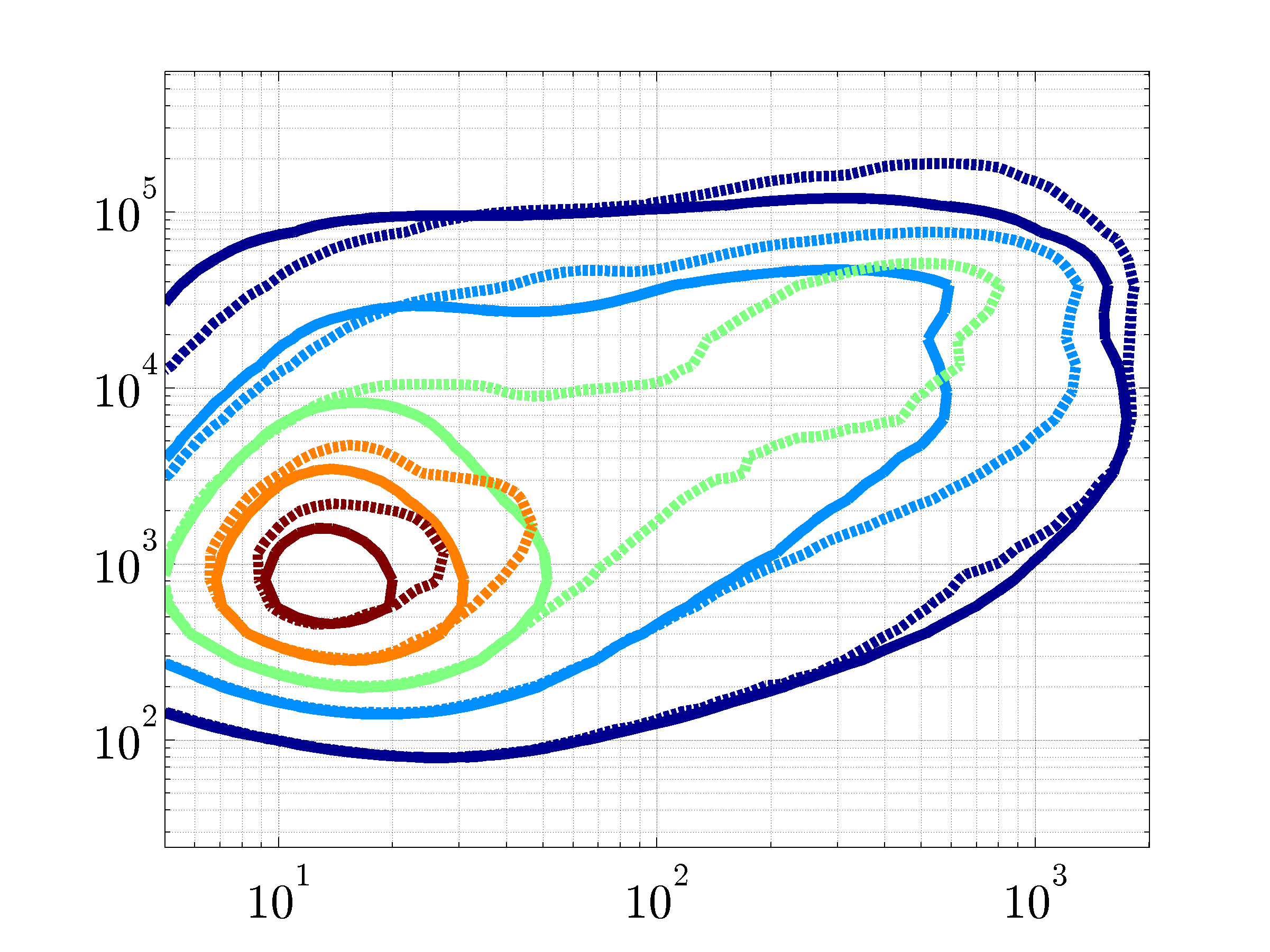}
    \label{fig.Euu_vs_yp_lxp_DNSb_Wr_R2003_match_uvw_uvr_y60_c0_100_Uc_s1_2_23_072613_c}}
    \\[0.2cm]
    $(a)$
    &
    \hskip-0.5cm
    $(b)$
    &
    \hskip-0.5cm
    $(c)$
    \end{tabular}
    \begin{tabular}{c}
    \\[-3.4cm]
    \begin{tabular}{c}
    \hskip-7cm
    \begin{turn}{90}
    $\lambda_x^+$
    \end{turn}
    \end{tabular}
    \\[1.5cm]
    \begin{tabular}{c}
    \hskip0.25cm
    $y^+$
    \hskip3.85cm
    $y^+$
    \hskip3.85cm
    $y^+$
    \end{tabular}
    \end{tabular}
    \\[-0.2cm]
    \begin{tabular}{ccc}
    \subfigure{\includegraphics[width=0.33\columnwidth]
    {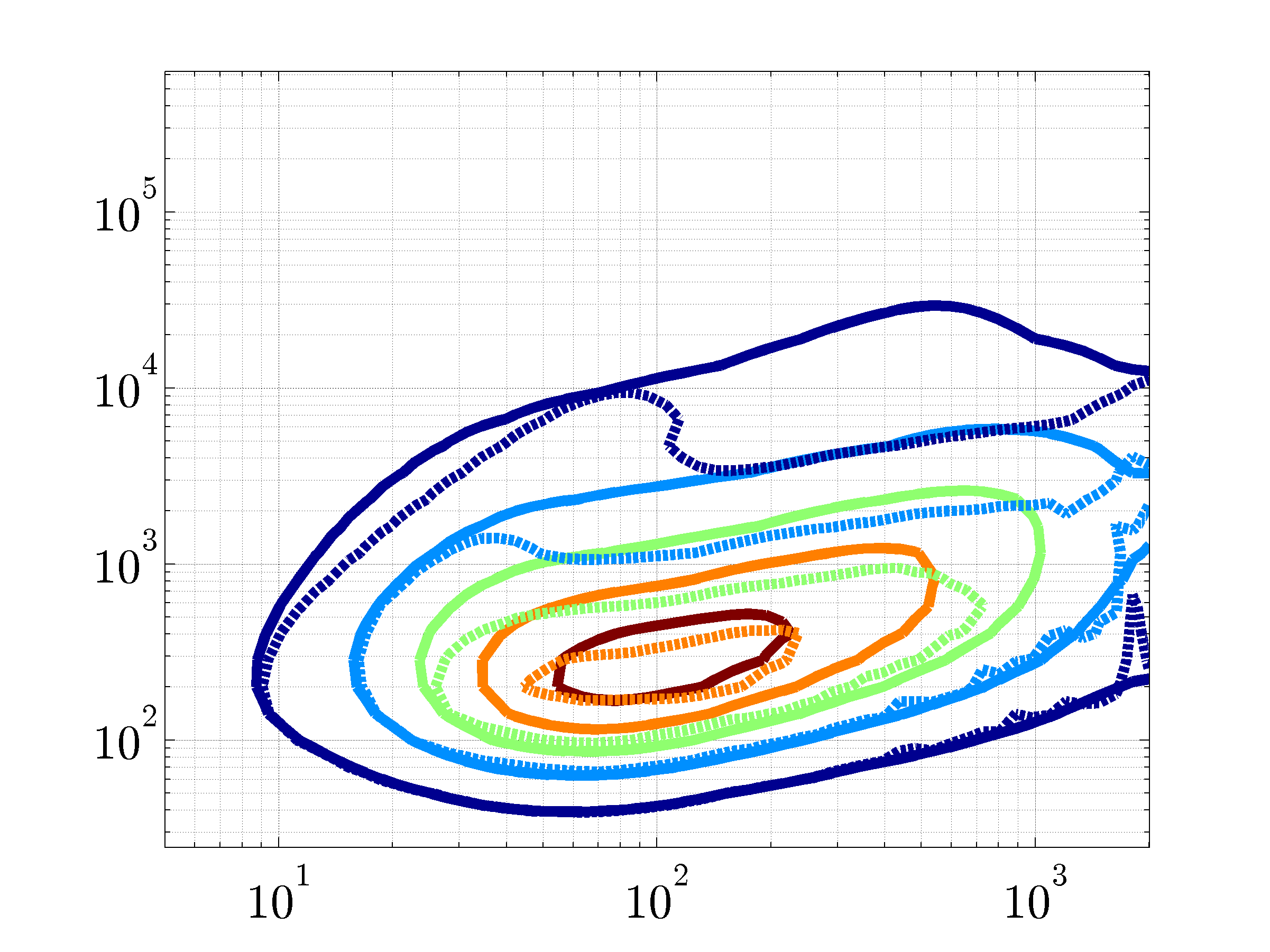}
    \label{fig.Evv_vs_yp_lxp_DNSb_Wr_R2003_match_uvw_uvr_y60_c0_100_Uc_s1_2_3_072613_c}}
    &
    \hskip-0.5cm
    \subfigure{\includegraphics[width=0.33\columnwidth]
    {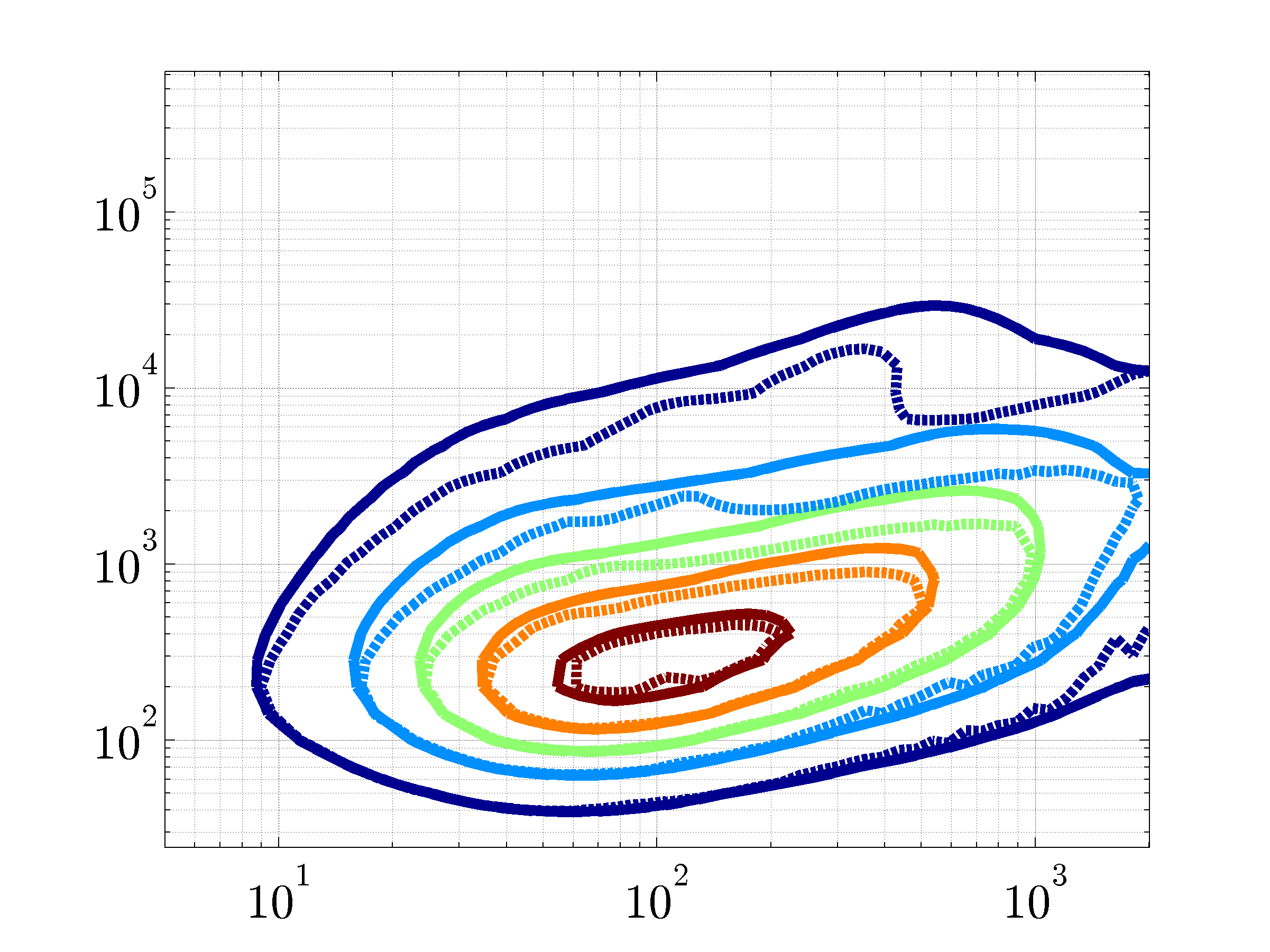}
    \label{fig.Evv_vs_yp_lxp_DNSb_Wr_R2003_match_uvw_uvr_y60_c0_100_Uc_s1_2_11_072613_c}}
    &
    \hskip-0.5cm
    \subfigure{\includegraphics[width=0.33\columnwidth]
    {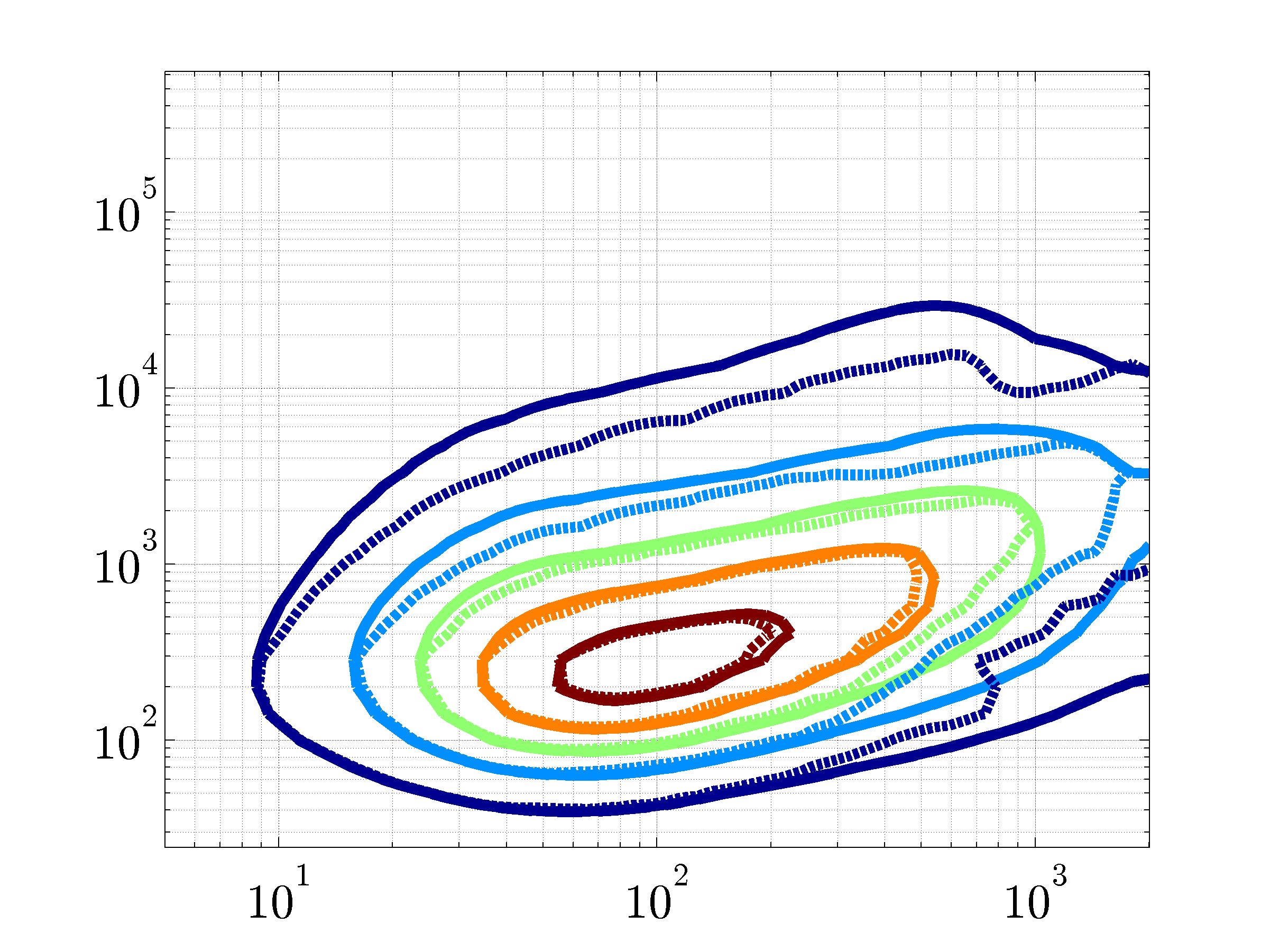}
    \label{fig.Evv_vs_yp_lxp_DNSb_Wr_R2003_match_uvw_uvr_y60_c0_100_Uc_s1_2_23_072613_c}}
    \\[0.2cm]
    $(d)$
    &
    \hskip-0.5cm
    $(e)$
    &
    \hskip-0.5cm
    $(f)$
    \end{tabular}
    \begin{tabular}{c}
    \\[-3.4cm]
    \begin{tabular}{c}
    \hskip-7cm
    \begin{turn}{90}
    $\lambda_x^+$
    \end{turn}
    \end{tabular}
    \\[1.5cm]
    \begin{tabular}{c}
    \hskip0.25cm
    $y^+$
    \hskip3.85cm
    $y^+$
    \hskip3.85cm
    $y^+$
    \end{tabular}
    \end{tabular}
    \\[-0.2cm]
    \begin{tabular}{ccc}
    \subfigure{\includegraphics[width=0.33\columnwidth]
    {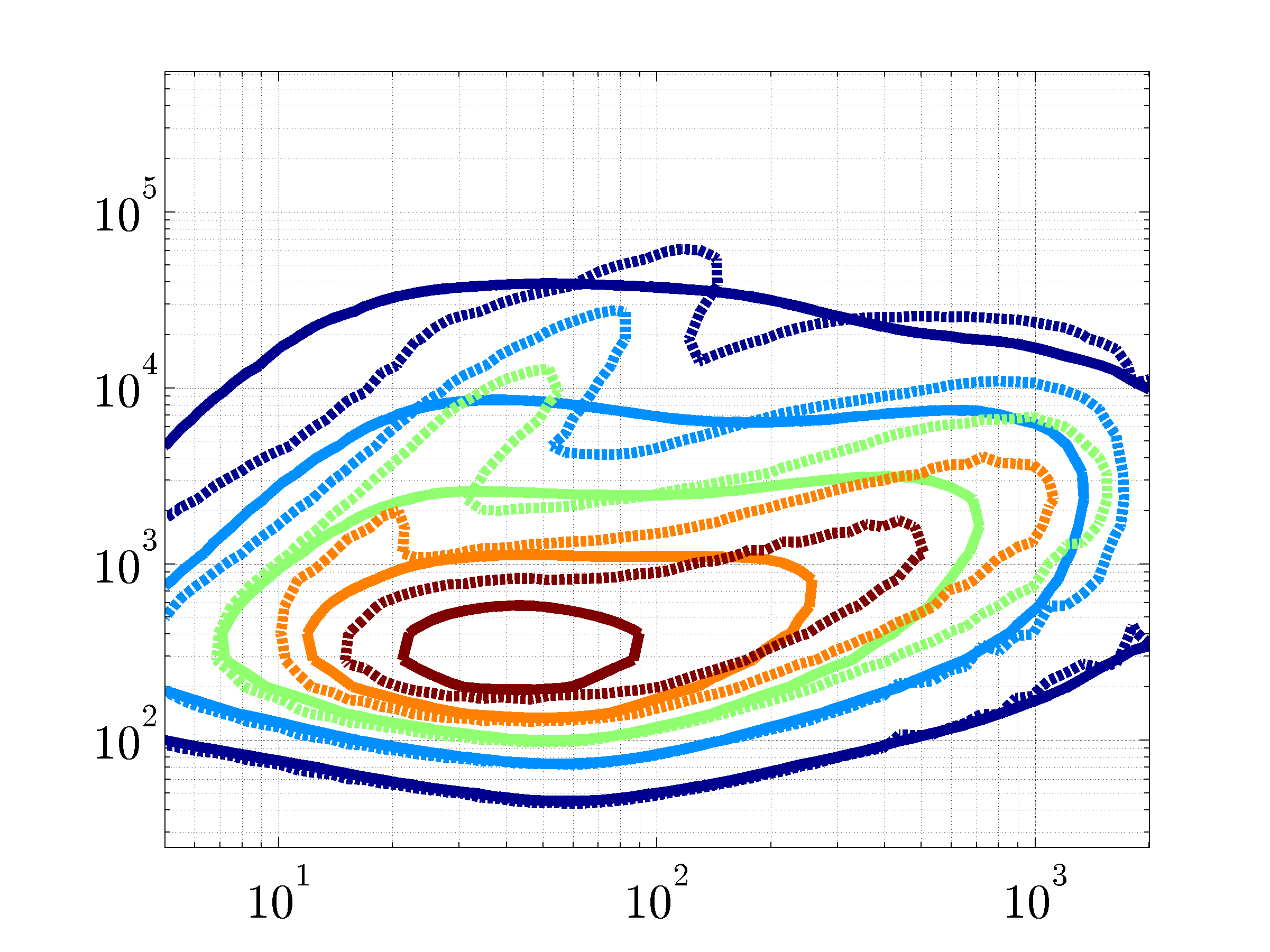}
    \label{fig.Eww_vs_yp_lxp_DNSb_Wr_R2003_match_uvw_uvr_y60_c0_100_Uc_s1_2_3_072613_c}}
    &
    \hskip-0.5cm
    \subfigure{\includegraphics[width=0.33\columnwidth]
    {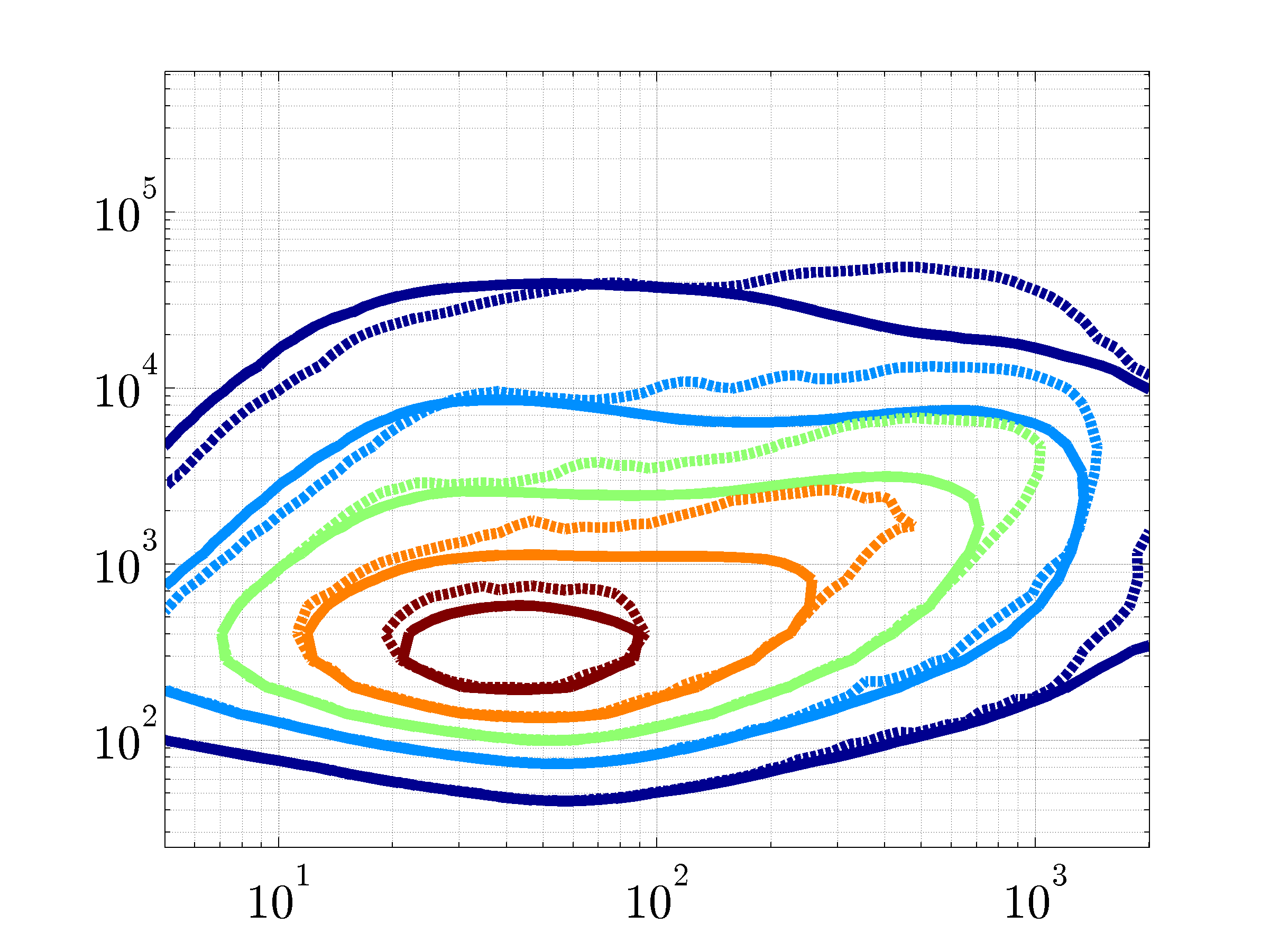}
    \label{fig.Eww_vs_yp_lxp_DNSb_Wr_R2003_match_uvw_uvr_y60_c0_100_Uc_s1_2_11_072613_c}}
    &
    \hskip-0.5cm
    \subfigure{\includegraphics[width=0.33\columnwidth]
    {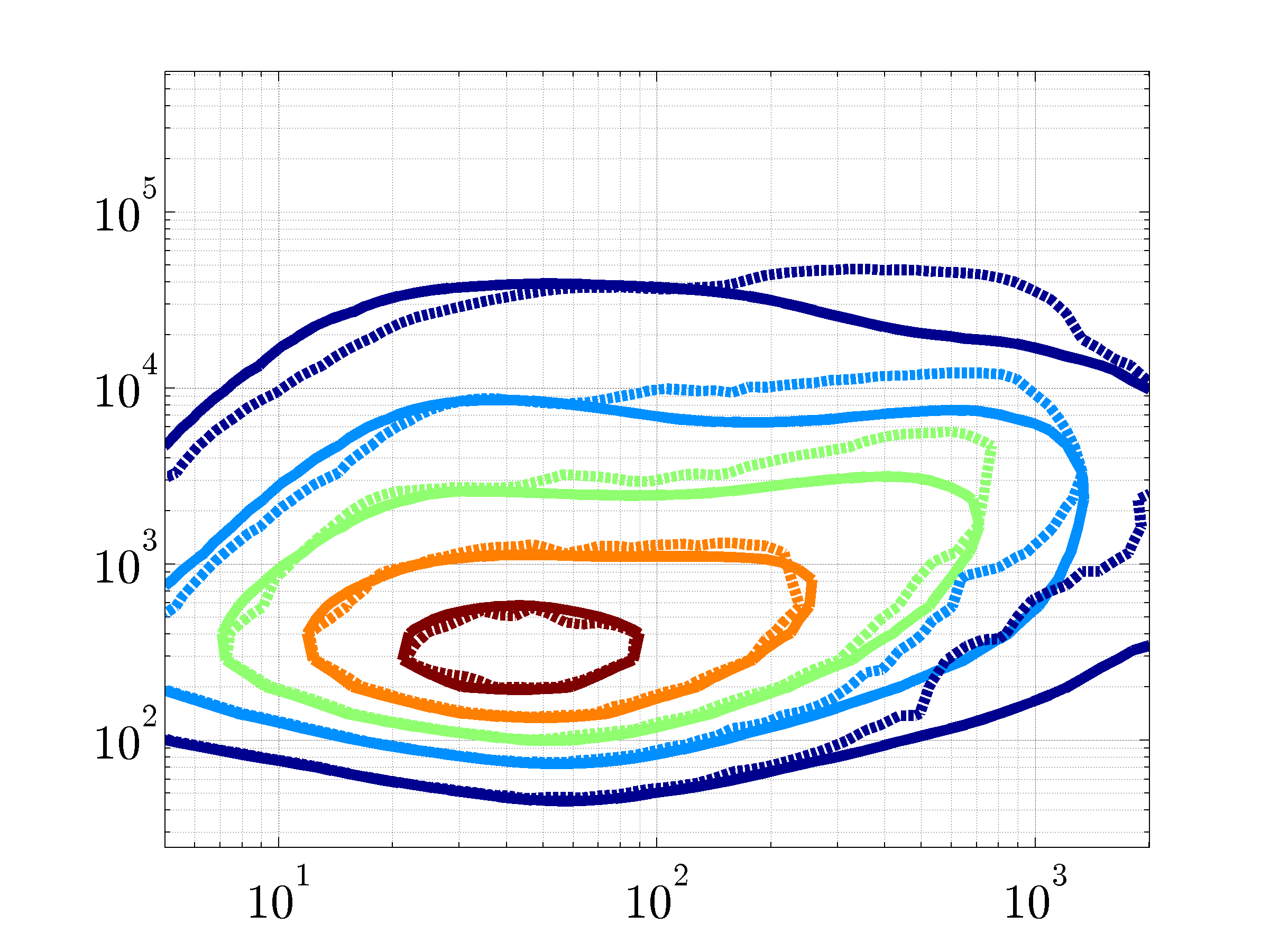}
    \label{fig.Eww_vs_yp_lxp_DNSb_Wr_R2003_match_uvw_uvr_y60_c0_100_Uc_s1_2_23_072613_c}}
    \\[0.2cm]
    $(g)$
    &
    \hskip-0.5cm
    $(h)$
    &
    \hskip-0.5cm
    $(i)$
    \end{tabular}
    \begin{tabular}{c}
    \\[-3.4cm]
    \begin{tabular}{c}
    \hskip-7cm
    \begin{turn}{90}
    $\lambda_x^+$
    \end{turn}
    \end{tabular}
    \\[1.5cm]
    \begin{tabular}{c}
    \hskip0.25cm
    $y^+$
    \hskip3.85cm
    $y^+$
    \hskip3.85cm
    $y^+$
    \end{tabular}
    \end{tabular}
    \\[-0.2cm]
    \begin{tabular}{ccc}
    \subfigure{\includegraphics[width=0.33\columnwidth]
    {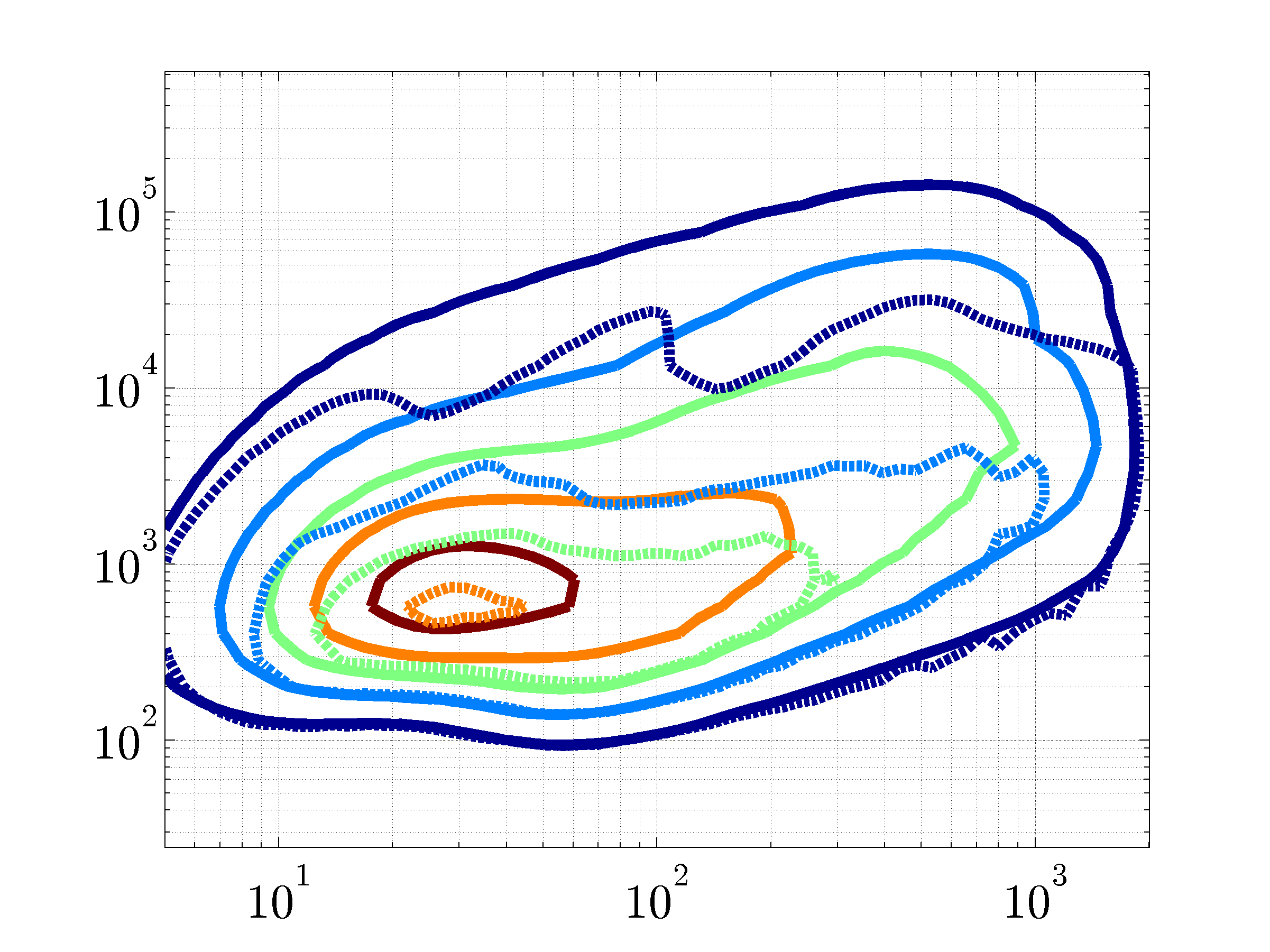}
    \label{fig.Euvr_vs_yp_lxp_DNSb_Wr_R2003_match_uvw_uvr_y60_c0_100_Uc_s1_2_3_072613_c}}
    &
    \hskip-0.5cm
    \subfigure{\includegraphics[width=0.33\columnwidth]
    {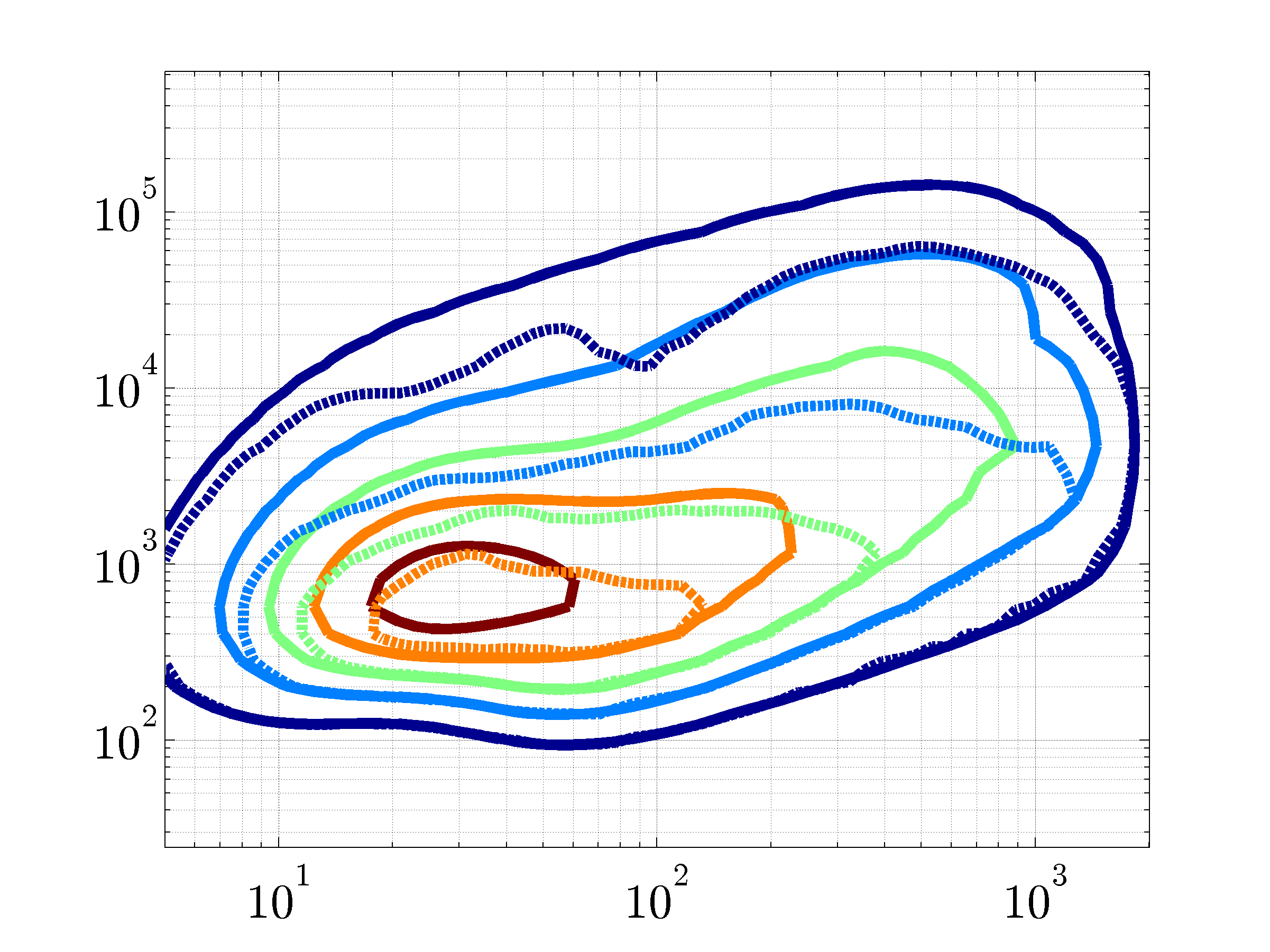}
    \label{fig.Euvr_vs_yp_lxp_DNSb_Wr_R2003_match_uvw_uvr_y60_c0_100_Uc_s1_2_11_072613_c}}
    &
    \hskip-0.5cm
    \subfigure{\includegraphics[width=0.33\columnwidth]
    {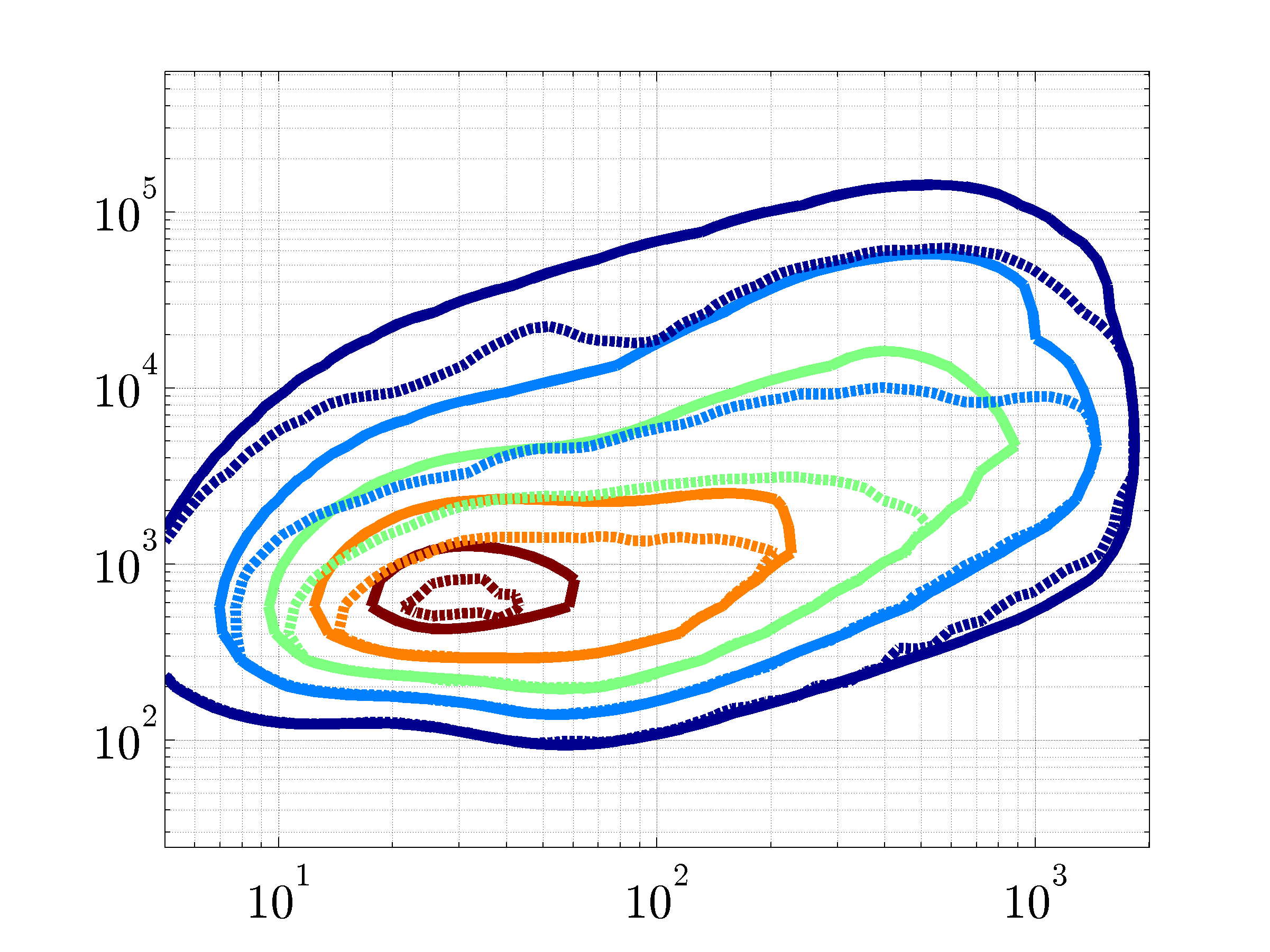}
    \label{fig.Euvr_vs_yp_lxp_DNSb_Wr_R2003_match_uvw_uvr_y60_c0_100_Uc_s1_2_23_072613_c}}
    \\[0.2cm]
    $(j)$
    &
    \hskip-0.5cm
    $(k)$
    &
    \hskip-0.5cm
    $(l)$
    \end{tabular}
    \begin{tabular}{c}
    \\[-3.4cm]
    \begin{tabular}{c}
    \hskip-7cm
    \begin{turn}{90}
    $\lambda_x^+$
    \end{turn}
    \end{tabular}
    \\[1.5cm]
    \begin{tabular}{c}
    \hskip0.25cm
    $y^+$
    \hskip3.85cm
    $y^+$
    \hskip3.85cm
    $y^+$
    \end{tabular}
    \end{tabular}
    \vspace{-0.3cm}
    \end{center}
    \captionsetup{justification=raggedright}
    \caption{    
    The solid contours are the time-averaged spectra from DNS~\cite{hoyjim06} for $Re_\tau = 2003$:
    (a)-(c) $E_{uu,\mathrm{DNS}}$, (d)-(f) $E_{vv,\mathrm{DNS}}$, (g)-(i) $E_{ww,\mathrm{DNS}}$, and (j)-(l) $-E_{uv,\mathrm{DNS}}$. 
    The dashed contours are the model-based spectra with the optimal weights using (a,d,g,j) $N = 2$, (b,e,h,k) $6$, and (c,f,i,l) $12$ resolvent modes per $\bkappa$ and $0 \leq c \leq U_{cl}$.
    The contours show $10 \%$ to $90 \%$ of the maximum in the DNS data with increments of $20 \%$. 
    }
    \label{fig.Euu-spectra-R2003}
    \end{figure}  
    
Problem~(\ref{eq.opt}) is solved using $N = 2$ to $12$ resolvent modes per $\bkappa$ and $c$. The channel symmetry around the center plane results in paired resolvent modes that are symmetric/anti-symmetric counterparts of each other~\cite{moashatromckJFM13}. For each pair of resolvent modes, the mode corresponding to the larger singular value is used. 
\moa{Even though the number of modes that are necessary for representing the spectra may vary for different wavenumber/speed combinations, considering a constant $N$ is sufficient to showcase the main trends.} 
\moa{We show that $N = 12$ yields a good agreement between the model-based spectra and the spectra obtained from DNS and that increasing $N$ beyond $12$ results in a diminishing return. The convergence analysis of the spectra as $N$ tends to infinity is beyond the scope of the present letter and a subject of ongoing research.}

Fig.~\ref{fig.Euu-spectra-R2003} compares the turbulent spectra from DNS~\cite{hoyjim06} (solid contours) and the spectra obtained using $N = 2$ (left column), $6$ (center column), and $12$ (right column) resolvent modes per $\bkappa$ and $c$ (dotted contours). The contours show $10 \%$ to $90 \%$ of the maximum in the DNS data with increments of $20 \%$. We see that even $2$ resolvent modes are sufficient to capture the general features of the turbulent spectra. The streamwise and spanwise spectra are better matched while the $90\%$ levels in the wall-normal and the $uv$ spectra are not captured. 

\begin{figure}
    \begin{center}
    \begin{tabular}{cc}
    \subfigure{\includegraphics[width=0.37\columnwidth]
    {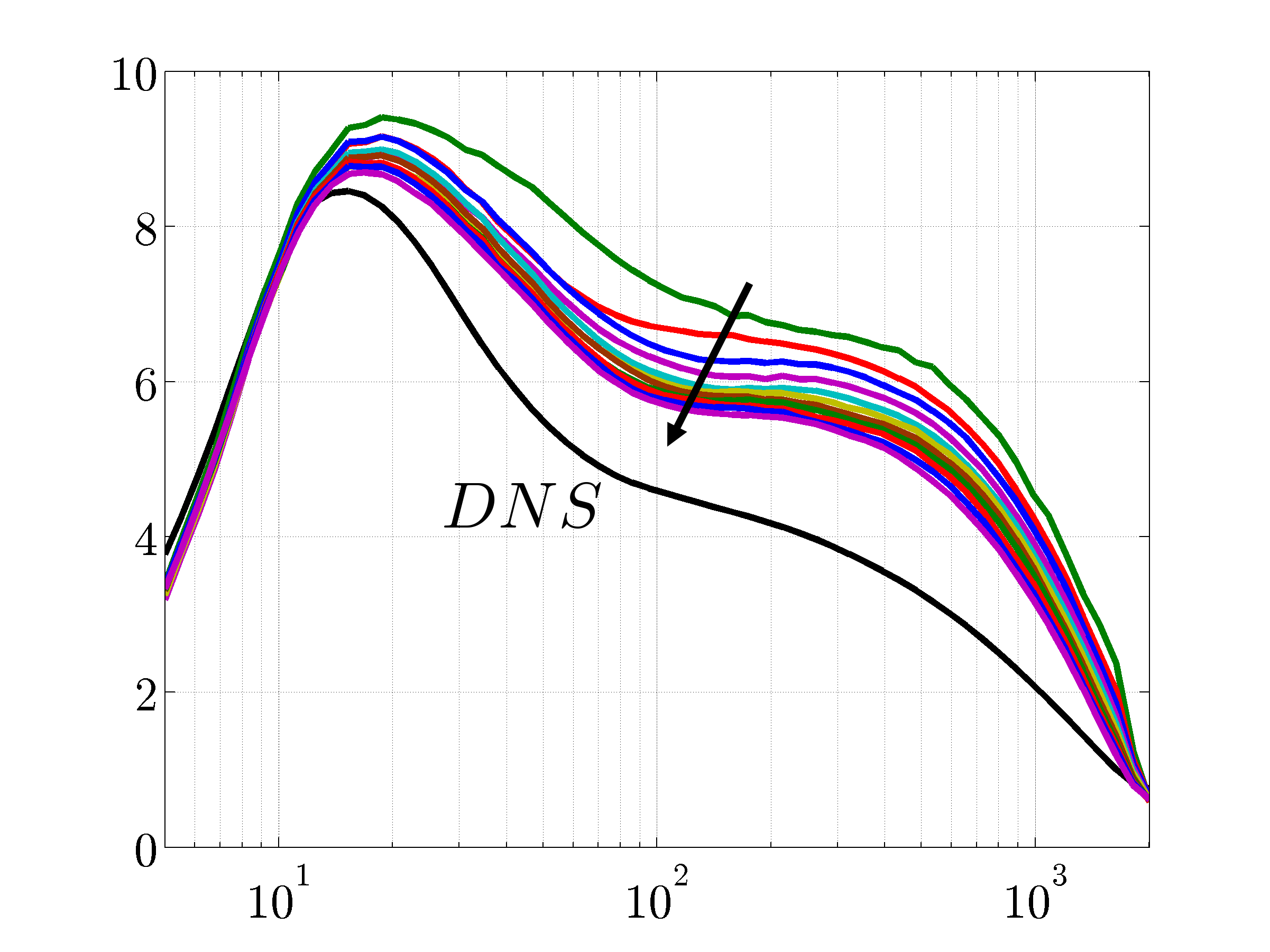}
    \label{fig.Euu_vs_yp_DNSk_R2003_match_uvw_uvr_y60_c0_100_Uc_072613_c}}
    &
    \subfigure{\includegraphics[width=0.37\columnwidth]
    {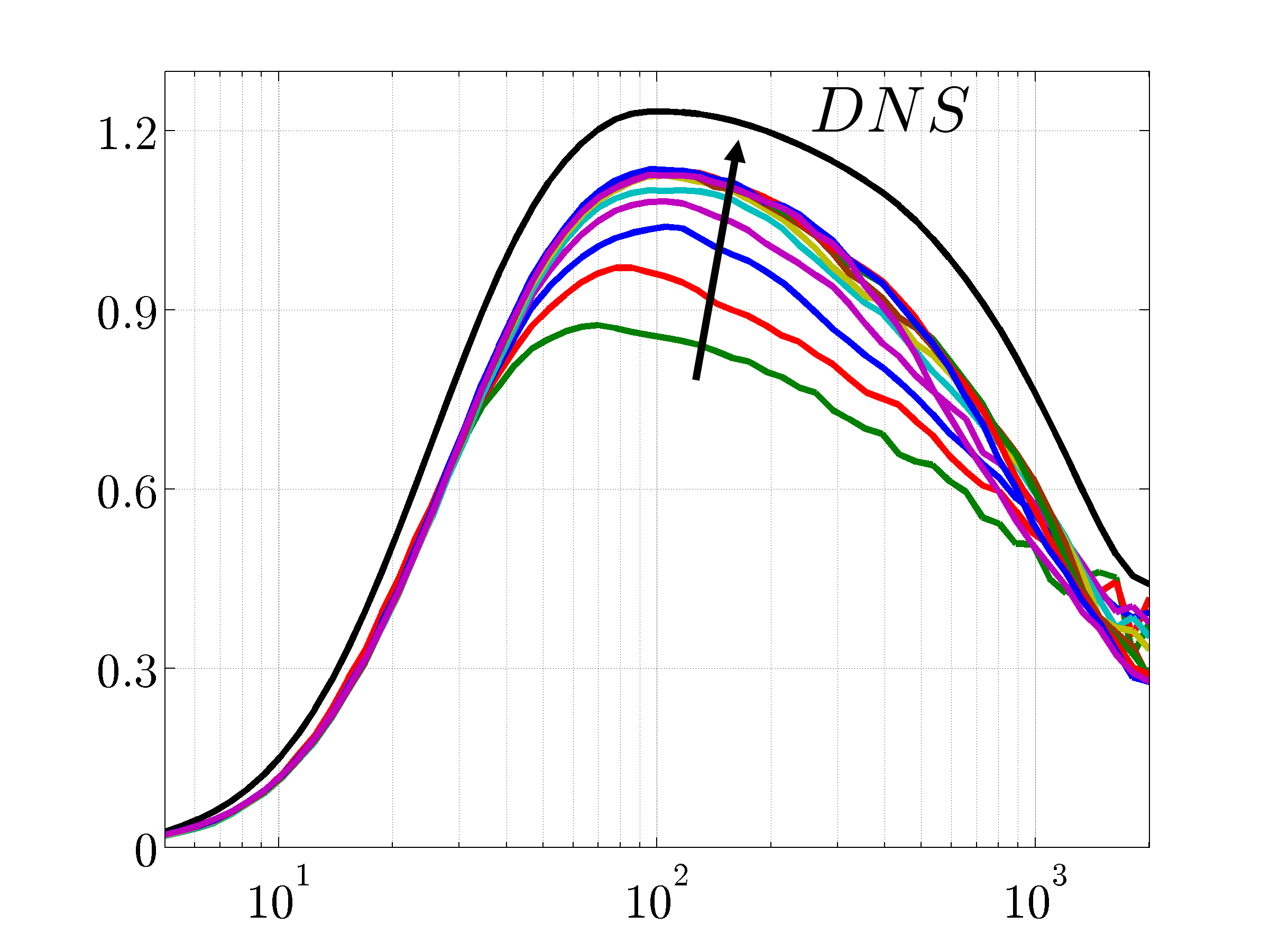}
    \label{fig.Evv_vs_yp_DNSk_R2003_match_uvw_uvr_y60_c0_100_Uc_072613_c}}
    \\[-2.55cm]
    \hskip-5.5cm
    \begin{turn}{90}
    $E_{uu}$
    \end{turn}
    &
    \hskip-5.5cm
    \begin{turn}{90}
    $E_{vv}$
    \end{turn}
    \\[1.6cm]
    $y^+$
    &
    $y^+$
    \\[0.05cm]
    $(a)$
    &
    $(b)$
    \\[-0.2cm]
    \subfigure{\includegraphics[width=0.37\columnwidth]
    {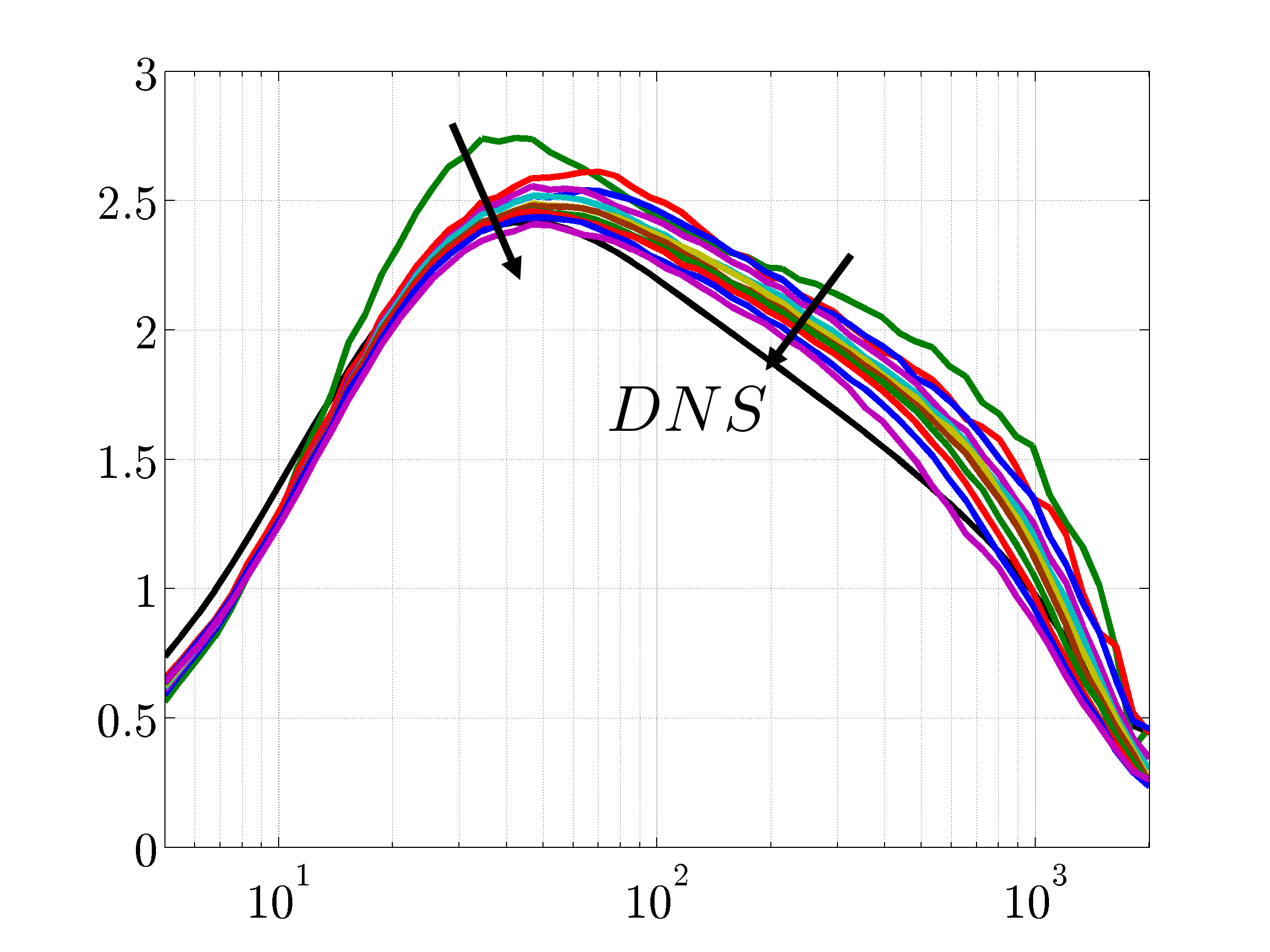}
    \label{fig.Eww_vs_yp_DNSk_R2003_match_uvw_uvr_y60_c0_100_Uc_072613_c}}
    &
    \subfigure{\includegraphics[width=0.37\columnwidth]
    {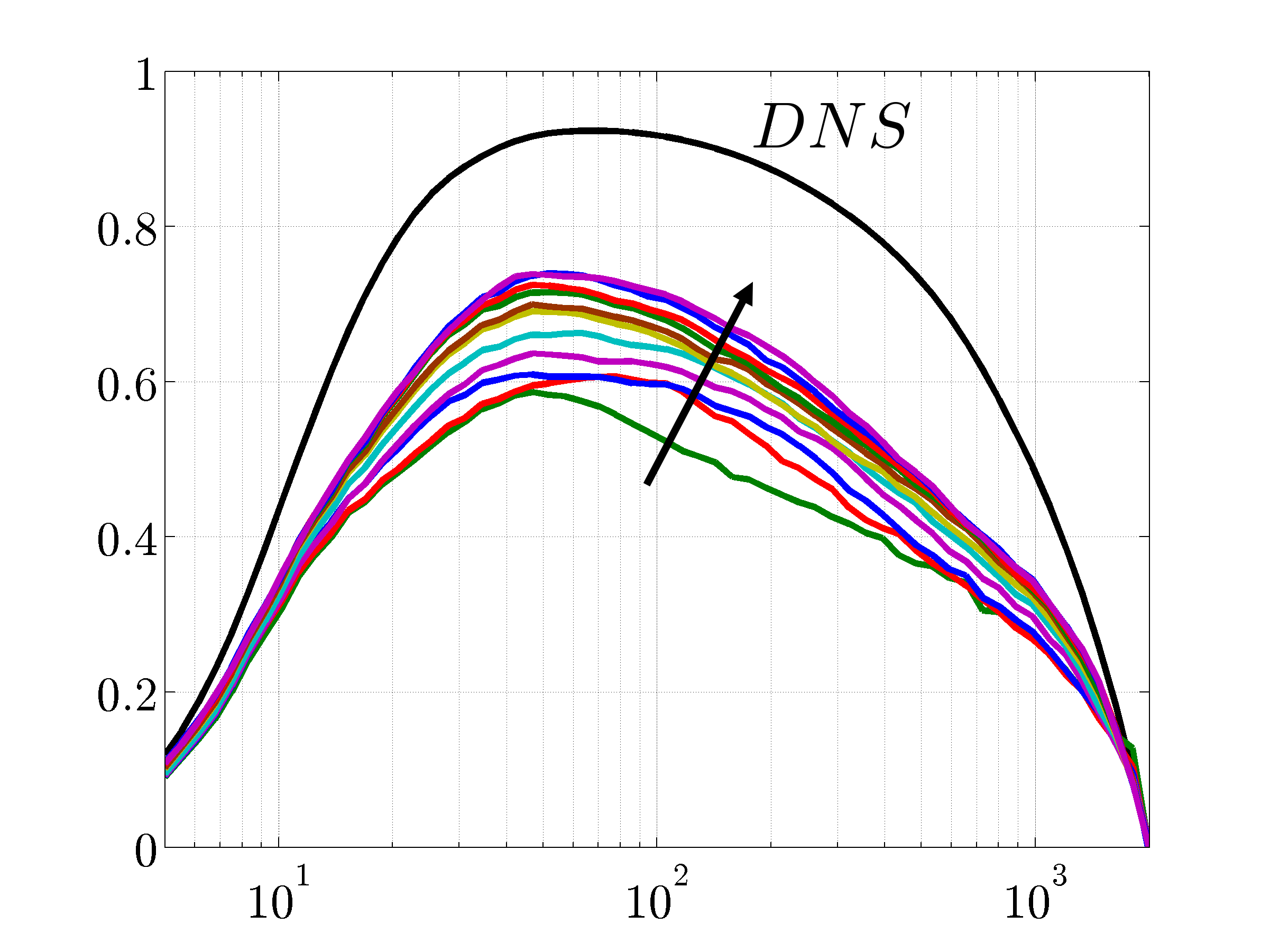}
    \label{fig.Euvr_vs_yp_DNSk_R2003_match_uvw_uvr_y60_c0_100_Uc_072613_c}}
    \\[-2.65cm]
    \hskip-5.5cm
    \begin{turn}{90}
    $~~E_{ww}$
    \end{turn}
    &
    \hskip-5.5cm
    \begin{turn}{90}
    $-E_{uv}$
    \end{turn}
    \\[1.5cm]
    $y^+$
    &
    $y^+$
    \\[0.05cm]
    $(c)$
    &
    $(d)$
    \end{tabular}
    \vspace{-0.3cm}
    \end{center}
    \captionsetup{justification=raggedright}
    \caption{    
    The black curves are the energy intensities from DNS~\cite{hoyjim06} for $Re_\tau = 2003$. 
    The colored curves are the model-based intensities with the optimal weights using $N = 2$ to $12$ resolvent modes per $\bkappa$ and $0 \leq c \leq U_{cl}$. Arrows show the direction of increasing $N$. 
    }
    \label{fig.Euu-R2003}
    \end{figure}

Using $6$ resolvent modes significantly improves matching of the wall-normal and spanwise spectra. The peaks of the streamwise and $uv$ spectra are matched more accurately even though the $90\%$ level in the $uv$ spectrum is still absent. Using $12$ resolvent modes results in close matching of the wall-normal and spanwise spectra and emergence of the $90\%$ level in the $uv$ spectrum. Notice that the spectra for small wavelengths ($\lambda_x^+ \lesssim 600$) are well-captured using $N = 12$ while representing the spectra for larger wavelengths requires more resolvent modes. The deviation error for the inner-scaled peak ($\lambda_x^+ = 700$, $\lambda_z^+ = 100$) and the outer-scaled peak ($\kappa_x = 0.6$, $\kappa_z = 6$) of the streamwise spectrum is respectively $22\%$ and $62\%$ using $N = 12$ resolvent modes. 
Fig.~\ref{fig.Euu-R2003} shows the energy intensities and the Reynolds stress obtained from DNS~\cite{hoyjim06} (black curves) and $N = 2$ to $12$ optimally weighted resolvent modes per $\bkappa$ and $c$ (colored curves). The arrows show the direction of increasing $N$. For $N = 12$, the deviation errors in the streamwise, wall-normal, and spanwise intensities and the Reynolds stress are respectively $20\%$, $17\%$, $6\%$, and $25\%$.

The present work shows that a small number of resolvent modes ($N N_c = 12 \times 100$) per wall-parallel wavenumber pair can simultaneously approximate the turbulent velocity spectra and the Reynolds stress co-spectrum in a turbulent channel. This number is less than $0.03\%$ of the number of degrees of freedom in DNS~\cite{hoyjim06} ($663 \times 7730$, where $633$ and $7730$ are respectively the number of wall-normal points and the number of temporal fields that are averaged to obtain the spectra). The actual reduction in the number of degrees of freedom is even larger since the optimal solution is sparse in the mode speed; for example, the weights corresponding to approximately $60\%$ and $80\%$ of the resolvent modes are zero for the inner- and outer-scaled peaks of the streamwise spectrum, respectively. In addition, we highlight that the resolvent-mode decomposition exhibits several important properties that are essential to predicting the behavior of wall-turbulence at high Reynolds numbers. We note that our approach does not capture the phase relationships between modes with different wall-parallel wavenumbers since this information is absent in the power spectra. 
\moa{In addition, while the computed weights represent the best fit to the spectra, they may not yield the exact solution of the velocity field.}
Our ongoing research is focused on determination of the weights by analyzing the nonlinear interaction of the resolvent modes.

\vskip0.5cm

The support of Air Force Office of Scientific Research under grants FA 9550-09-1-0701 (P.M. Rengasamy Ponnappan) and FA 9550-12-1-0469 (P.M. Doug Smith) is gratefully acknowledged.

\bibliographystyle{aiaa}
\bibliography{../bib/couette,../bib/mj-complete-bib,../bib/periodic,../bib/covariance,../bib/control-pde,../bib/ref-added-rm}

\begin{thebibliography}{10}
\newcommand{\enquote}[1]{``#1''}

\bibitem{smimckmar11}
Smits, A.~J., McKeon, B.~J., and Marusic, I., \enquote{High-{R}eynolds number
  wall turbulence,} {\em Annu. Rev. Fluid Mech.\/}, Vol.~43, 2011,
  pp.~353--375.

\bibitem{berhollum93}
Berkooz, G., Holmes, P., and Lumley, J.~L., \enquote{The proper orthogonal
  decomposition in the analysis of turbulent flows,} {\em Annu. Rev. Fluid
  Mech.\/}, Vol.~25, 1993, pp.~539--575.

\bibitem{row05}
Rowley, C.~W., \enquote{Model reduction for fluids using balanced proper
  orthogonal decomposition,} {\em Int. J. Bifurcation Chaos\/}, Vol.~15, No.~3,
  2005, pp.~997--1013.

\bibitem{sch10}
Schmid, P.~J., \enquote{Dynamic mode decomposition of numerical and
  experimental data,} {\em J. Fluid Mech.\/}, Vol.~656, 2010, pp.~5--28.

\bibitem{tum11}
Tumin, A., \enquote{The biorthogonal eigenfunction system of linear stability
  equations: {A} survey of applications to receptivity problems and to analysis
  of experimental and computational results,} Presented at AIAA Fluid Dyn.
  Conf. Exhib., 41st, Honolulu, AIAA Pap. 2011-3244.

\bibitem{mez13}
Mezi{\'c}, I., \enquote{Analysis of fluid flows via spectral properties of the
  {K}oopman operator,} {\em Annu. Rev. Fluid Mech.\/}, Vol.~45, 2013,
  pp.~357--378.

\bibitem{mcksha10}
McKeon, B.~J. and Sharma, A.~S., \enquote{A critical-layer framework for
  turbulent pipe flow,} {\em J. Fluid Mech.\/}, Vol.~658, 2010, pp.~336--382.

\bibitem{mckshajac13}
McKeon, B.~J., Sharma, A.~S., and Jacobi, I., \enquote{{Experimental
  manipulation of wall turbulence: A systems approach},} {\em Phys. Fluids\/},
  Vol.~25, 2013, pp.~031301.

\bibitem{shamck13}
Sharma, A.~S. and McKeon, B.~J., \enquote{On coherent structure in wall
  turbulence,} {\em J. Fluid Mech.\/}, Vol.~728, 2013, pp.~196--238.

\bibitem{moashatromckJFM13}
Moarref, R., Sharma, A.~S., Tropp, J.~A., and McKeon, B.~J.,
  \enquote{Model-based scaling of the streamwise energy density in
  high-{R}eynolds number turbulent channels,} {\em J. Fluid Mech.\/}, Vol.~734,
  2013, pp.~275--316.

\bibitem{hoyjim06}
Hoyas, S. and Jim{\'e}nez, J., \enquote{Scaling of the velocity fluctuations in
  turbulent channels up to ${R}e_{\tau} = 2003$,} {\em Phys. Fluids\/},
  Vol.~18, No.~1, 2006, pp.~011702.

\bibitem{col56}
Coles, D.~E., \enquote{The law of the wake in the turbulent boundary layer,}
  {\em J. Fluid Mech.\/}, Vol.~1, 1956, pp.~191--226.

\bibitem{lehguamck11}
LeHew, J., Guala, M., and McKeon, B.~J., \enquote{A study of the
  three-dimensional spectral energy distribution in a zero pressure gradient
  turbulent boundary layer,} {\em Exp. Fluids\/}, Vol.~51, 2011, pp.~997--1012.

\bibitem{moashatromck13-AIAA}
Moarref, R., Sharma, A.~S., Tropp, J.~A., and McKeon, B.~J., \enquote{On
  effectiveness of a rank-1 model of turbulent channels for representing the
  velocity spectra,} {\em 43rd AIAA Fluid Dyn. Conf., 2013-2480\/}, 2013.

\bibitem{huapal10}
Huang, Y. and Palomar, D.~P., \enquote{Rank-constrained separable semidefinite
  programming with applications to optimal beamforming,} {\em IEEE T. Signal
  Proces.\/}, Vol.~58, No.~2, 2010, pp.~664--678.

\bibitem{cvx}
{CVX Research, Inc.}, \enquote{{CVX}: Matlab Software for Disciplined Convex
  Programming, version 2.0 beta,} \url{http://cvxr.com/cvx}, Sept. 2012.

\end{thebibliography}

\end{document}